\documentstyle[preprint,aps,version2]{revtex}
\begin{document}
\draft
%
\begin{title}
Spin-Orbit Scattering and Pair Breaking in a Structurally\\
Disordered Copper-Oxide Layer
\end{title}
\author{N.~E.~Bonesteel\cite{currentaddress}}
\begin{instit}
Theoretische Physik, Eidgen\"ossische Technische
Hochschule-H\"onggerberg,\\ CH-8093 Z\"urich, Switzerland
\end{instit}
\receipt{    }
\begin{abstract}
To leading order in displacement size, the scattering of electrons in
a Cu-O plane from O displacements perpendicular to that plane is due
to spin-orbit coupling.  This scattering is investigated with the
following results: (1) As a consequence of time-reversal symmetry,
spin fluctuations, which can strongly enhance scattering from a spin
impurity, do not enhance spin-orbit scattering; and (2) for a
superconductor with a $d_{x^2-y^2}$ gap function, pair-breaking from
spin-orbit scattering can be strong, particularly in a structurally
disordered phase in which locally CuO$_6$ octahedra tilt as in the
orthorhombic phase of La$_2$CuO$_4$, but globally the average
structure is tetragonal.  These results are discussed in the context
of the (La,Nd)-(Sr,Ba)-Cu-O system where certain structural
transitions are observed to suppress superconductivity.
\end{abstract}
\narrowtext
%
The microscopic origin of high transition temperatures in the cuprate
superconductors is still unknown, even after six years of intense
study.  One school of thought holds that superconductivity in these
materials arises from the exchange of nearly antiferromagnetic spin
fluctuations.  If this is the case then it is almost certain that
Cooper pairs form with $d_{x^2-y^2}$ symmetry \cite{dwave}.

There is now a great deal of experimental evidence which shows an
intriguing interplay between small changes in lattice structure and
superconductivity in the La-based cuprates
\cite{axe,crawford,buchner}.  This interplay was first observed in the
La$_{2-x}$Ba$_x$CuO$_4$ system which, when $x \simeq 0.12$, undergoes
two structural phase transitions \cite{axe}.  The first transition is
from an undistorted high-temperature tetragonal phase into a
low-temperature orthorhombic (LTO) phase.  In the LTO phase the
CuO$_6$ octahedra making up each Cu-O layer tilt in a staggered
fashion about the $(110)$ axis.  The second transition is from the LTO
phase into a low-temperature tetragonal (LTT) phase in which, on
average, the CuO$_6$ octahedra tilt first about the $(100)$ and then
the $(010)$ axes in successive layers.  In this new phase
superconductivity appears to be completely destroyed \cite{axe}, and
recent experiments on the La$_{2-x-y}$Nd$_y$Sr$_x$CuO$_4$ system show
a similar correlation between unusual low temperature structural
phases ({\it i.e.}, the LTT phase, and another phase with space group
{\it Pccn}, intermediate between the LTO and LTT phases) and
suppression of superconductivity
\cite{crawford,buchner}.

One possible explanation for these experiments is that this
suppression of superconductivity is due to pair breaking
\cite{abrikosov,maki}.  It is a well-known characteristic of
unconventional pairing, such as $d$-wave, that the superconducting
transition temperature, $T_c$, is sensitive to elastic impurity
scattering \cite{pairbreaking}.  Because the LTT and {\it Pccn} phases
are stabilized by random substitution of Nd or Ba ions for La, it is
likely that these phases contain more structural disorder than the LTO
phase.  If so, then elastic scattering of electrons from this disorder
may be responsible for the observed suppression of superconductivity.

The tilting of a CuO$_6$ octahedon in a given Cu-O plane causes O ions
to be displaced out of that plane. In what follows a `one-band'
Hamiltonian is used to describe the coupling of electrons to these
displacements
\cite{brz}:
\begin{equation}
H_0 = -t \sum_{\scriptstyle i,j \atop{\scriptstyle{\alpha\beta}}}
c^\dagger_{i\alpha} \left((1 - \rho \theta^2_{ij})
\delta_{\alpha\beta}
+ i\nu\theta_{ij}{\bf \hat
\eta}_{ij}\cdot{\vec\sigma}_{\alpha\beta}\right)
c^{\phantom{\dagger}}_{j\beta}\label{h0}.
\end{equation}
\narrowtext\noindent
The index $i$ labels Cu sites on a two-dimensional square lattice with
$N$ sites, $c^\dagger_{i\alpha}$ is the creation operator for an
electron with spin $\alpha$ at site $i$, and $\theta_{ij}$ is the
angle between the Cu-O plane and the bond made by the Cu ion at site
$i$ and the O ion between sites $i$ and $j$.  Recent microscopic
calculations have found that $\hat\eta_{i,i+{\hat{x}}}
\simeq {\hat y}$ and $\hat\eta_{i,i+{\hat{y}}} \simeq -{\hat x}$
\cite{neb,shekhtman,koshibae}.  Hamiltonian (\ref{h0}) describes two
distinct electron-lattice couplings: (i) the spin-independent
$\theta^2$ coupling which arises from the quadratic modification of
the Cu-O bond lengths in the presence of an O displacement; and (ii)
the linear in $\theta$ coupling which occurs through spin-orbit
\cite{brz}.  At half-filling (one electron per site) (ii) is
responsible for the anisotropic Dzyaloshinki-Moriya corrections to
the otherwise isotropic superexchange interaction between Cu spins
\cite{moriya}.  The size of these corrections are known from
experiment \cite{wf} and can be used to estimate $\nu$ \cite{neb}.
The parameter values used here are $t\sim 400$ meV \cite{hyb}, $\nu
\sim 0.2$ \cite{neb,wf}, and $\rho$ is expected to be of order 1.

For a coherent tilting distortion
\begin{eqnarray}
\theta_{i,i+{\hat x}} &=&\theta_0
\sin\chi \exp{(i{\bf Q}\cdot{\bf r}_i)},\nonumber\\
\theta_{i,i+{\hat y}} &=&
\theta_0 \cos\chi \exp{(i{\bf Q}\cdot{\bf r}_i)},
\end{eqnarray}
where ${\bf Q}\equiv(\pi,\pi)$, and where $\chi = \pi/4$ in the LTO
phase, $\chi = 0$ in the LTT phase, and $0 < \chi < \pi/4$ in the {\it
Pccn} phase.  These coherent distortions cause Bragg scattering of
electrons through the spin-orbit coupling term in (\ref{h0}).  For
Bragg scattering it is possible to rediagonalize (\ref{h0}) so that
there is no scattering; however, a random component to $\theta_{ij}$
will give rise to ergodic scattering.

Before proceeding it is useful to contrast spin-orbit scattering as
described by (\ref{h0}) with spin-impurity scattering as described by
the interaction Hamiltonian
\begin{equation}
H_{spin} = J\sum_i {\bf S}_i \cdot c_i^\dagger {\vec \sigma}
c_i^{\phantom{\dagger}}.\label{hspin}
\end{equation}
Electrons will scatter elastically from a random displacement field
$\theta_{ij}$ as well as a random spin configuration ${\bf S}_i$
through the couplings in (\ref{h0}) and (\ref{hspin}).  Although both
scattering processes involve spin, there is an important difference:
Spin impurities are not time-reversal invariant perturbations ($\bf S
\rightarrow -{\bf S}$ under time reversal) while spin-orbit impurities, {\it
i.e.}, O displacements, are ($\theta \rightarrow \theta$ under time
reversal).  One well known consequence of this difference is that
spin-impurity scattering is pair breaking for a conventional $s$-wave
superconductor \cite{abrikosov}, but spin-orbit scattering is not
(Anderson's theorem) \cite{pwa}.

Another consequence of time-reversal symmetry appears when one
considers the possible spin-fluctuation enhancement of the scattering
vertex for spin-impurity and spin-orbit scattering. A spin impurity
embedded in an electron fluid polarizes the spins which surround it.
If this fluid is characterized by strong spin fluctuations, the
polarized region can be quite large.  Quasiparticles then scatter from
the impurity spin together with its polarization cloud, and this leads
to enhanced scattering at the characteristic spin fluctuation wave
vectors.  Because both spin-orbit and spin-impurity scattering involve
a spin flip, it is natural to ask if spin-orbit scattering can be
similarly enhanced by spin fluctuations.

To answer this question, consider adding a Hubbard $U$ interaction,
($H_{\rm Hub.} = U\sum_i n_{i\uparrow}n_{i\downarrow}$), to (\ref{h0})
and (\ref{hspin}).  For both spin-orbit and spin-impurity scattering
the renormalized scattering vertex can be written
$\Gamma_{\alpha\beta}({\bf k},{\bf k}+{\bf q}) =
\vec\Lambda_{{\bf k},{\bf k+q}} \cdot \vec
\sigma_{\alpha\beta}$.  Figure \ref{diagrams} shows the diagrammatic
equation for $\Gamma$ where $H_{Hub.}$ is treated in the random-phase
approximation (RPA).  The corresponding self-consistent equation for
$\vec\Lambda^{(RPA)}$ is
\FL
\begin{equation}
\vec\Lambda^{(RPA)}_{{\bf k},{\bf k+q}} =
\vec\Lambda^{(0)}_{{\bf k},\bf {k+q}} + U
\int {d^2 p\over{(2\pi)^2}} {f_{{\bf p}+{\bf q}} -
f_{\bf p}\over{\epsilon_{{\bf p}+{\bf q}}-\epsilon_{\bf p}}}
\vec\Lambda^{(RPA)}_{{\bf p},{\bf p}+{\bf q}}\label{vertex}.
\end{equation}
Here $\epsilon_{\bf q} = -2t(\cos q_x+\cos q_y) - \mu$ where $\mu$ is
the chemical potential, and $f_p \equiv f(\epsilon_p)$ is the Fermi
function.  Time-reversal symmetry requires that $\vec\Lambda_{{\bf
k},{\bf k}^\prime}=\pm
\vec\Lambda_{-{\bf k},-{\bf k}^\prime}$ with the $+$ and $-$ signs holding for
spin-impurity and spin-orbit scattering, respectively.  Because of
this difference the solution to (\ref{vertex}) is
$\vec\Lambda^{(RPA)}_{{\bf k},{\bf k}^\prime} = (1-U\chi_0({\bf
k}-{\bf k}^\prime))^{-1} {\vec\Lambda^{(0)}_{{\bf k},{\bf k}^\prime}}$
for spin-impurity scattering, where $\chi_0({\bf q})$ is the static
spin susceptibility for non-interacting electrons, and
$\vec\Lambda^{(RPA)}_{{\bf k},{\bf k}^\prime} =
{{\vec\Lambda^{(0)}_{{\bf k},{\bf k}^\prime}}}$ for spin-orbit
scattering.  Thus, as a consequence of time-reversal symmetry, the
ladder diagrams shown in Fig.~\ref{diagrams}, which enhance
spin-impurity scattering when $1-U\chi_0({\bf k}-{\bf k}^\prime)$ is
small, do not enhance spin-orbit scattering.

Next we proceed with the conventional pair-breaking analysis
\cite{abrikosov}, which begins with the linearized Gor'kov-Dyson
equations in the Matsubara formalism
\widetext
\begin{equation}
\Delta_{{\bf k},n} = -\pi T\sum_m^{|\omega_m| < \omega_{\rm SF}}
\int {d\theta_{{\bf k}^\prime}\over{2\pi}} N(\theta_{{\bf k}^\prime})
{\Delta_{{\bf k}^\prime,m}\over{|Z_{{\bf k}^\prime,m}||\omega_m|}}
\left(V_{{\bf k},{\bf k}^\prime}-{1\over T} (|v^{(-)}_{{\bf k},
{\bf k}^\prime}|^2+|w_{{\bf k},{\bf k}^\prime}|^2)
\delta_{m,n}\right),
\label{eliash1}
\end{equation}
\begin{equation}
\Sigma_{{\bf k},n} =  i\omega_n(1-Z_{{\bf k},n})
= -i\pi{\rm sgn}(\omega_n) \int {d\theta_{{\bf k}^\prime}\over{2\pi}}
N(\theta_{{\bf k}^\prime})(|v^{(+)}_{{\bf k},{\bf
k}^\prime}|^2+|w_{{\bf k},{\bf k}^\prime}|^2),
\label{eliash2}
\end{equation}
\narrowtext\noindent
where $\Delta_{{\bf k},n}$ and $\Sigma_{{\bf k},n}$ are the anomalous
and normal self energies, $\omega_n = (2n+1)\pi T$ is the $n$th
Matsubara frequency, the Fermi surface is parameterized by the angle
$\theta_{\bf k}$, and $N(\theta_{\bf k})$ is the local density of
states.

The phenomenological effective pairing interaction in (\ref{eliash1})
is taken to be
\begin{equation}
V_{{\bf k},{\bf k}^\prime} = -\lambda\phi_d({\bf k})
\phi_d({\bf k}^\prime)\label{pi}
\end{equation}
where $\phi_d({\bf k})=A(\cos k_x-\cos k_y)$ with $A=(\int
(d\theta_{\bf k}/2\pi)N(\theta_{\bf k}) (\cos k_x - \cos
k_y)^2)^{-1/2}$.  For $\lambda > 0$ this interaction is attractive in
the $d_{x^2-y^2}$ channel.  The sum over Matsubara frequencies in
(\ref{eliash1}) must be cut off for large frequencies.  Within the
spin-fluctuation model the cutoff $\omega_{\rm SF}$ should be viewed
as a characteristic spin-fluctuation frequency.  The critical
temperature $T_c$ is determined by finding the temperature at which
(\ref{eliash1}) and (\ref{eliash2}) have a nontrivial solution.

The functions $|v^{(\pm)}_{{\bf k},{\bf k}^\prime}|^2$ and $|w_{{\bf
k},{\bf k}^\prime}|^2$ in (\ref{eliash1}) and (\ref{eliash2}) are the
scattering matrix elements coming from the spin-orbit and
spin-independent couplings in (\ref{h0}), respectively.  To leading
order in $\theta$
\FL
\begin{equation}
|v^{(\pm)}_{{\bf k},{\bf k}^\prime}|^2 =
4t^2\nu^2 \sum_{a,b\in\{x,y\} }C_{ab}({\bf k}\pm {\bf k}^\prime)
\sin({k_a\pm k_a^\prime\over{2}})\sin({k_b\pm k_b^\prime
\over{2}})
\end{equation}
and
\FL
\begin{equation}
|w_{{\bf k},{\bf k}^\prime}|^2 = 4t^2\rho^2\sum_{a,b\in\{x,y\}}
F_{ab}({\bf k}-{\bf k}^\prime)\cos({k_a -
k_a^\prime\over{2}})\cos({k_b- k_b^\prime\over{2}})
\end{equation}
with
\FL
\begin{eqnarray}
C_{ab}({\bf q}) &=& \langle\theta_{{\bf q},a}
\theta_{-{\bf q},b}\rangle\label{ccf}
\\ F_{ab}({\bf q}) &=& \int
{d^2p\over{(2\pi)^2}}\int {d^2p^\prime\over{(2\pi)^2}}
\langle\theta_{{\bf q}+{\bf p},a}\theta_{-{\bf
p},a}\theta_{-{\bf q}+{\bf p}^\prime,b}
\theta_{-{\bf p}^\prime,b}\rangle \label{fcf}
\end{eqnarray}
where $\langle...\rangle$ denotes an average over disorder, and
$\theta_{{\bf q},a} = 1/N\sum_i \exp(i{\bf q}\cdot{\bf
r}_i)\theta_{i,i+{\hat a}}$.  Although spin-orbit scattering enters
the equations for the anomalous and normal self energies differently
because of the spin flip, for even-parity singlet pairing ${\bf
k}^\prime$ can be replaced by $-{\bf k}^\prime$ in (\ref{eliash1}).
Accordingly the $(\pm)$ superscript is suppressed in what follows.

Assuming the gap function can be factorized as $\Delta_{{\bf
k},m}/Z_{{\bf k},m} =
\phi_d(k) \tilde\Delta_m$ then (\ref{eliash1}) and (\ref{eliash2}) can
be combined to yield
\begin{equation}
{\tilde\Delta}_n
= \pi T \sum_m^{|\omega_m|<\omega_{\rm SF}}
{{\tilde\Delta}_m \over{|\omega_m|}}
\left(\lambda - \delta_{m,n}{1\over{\pi T\tau_{\rm pb}}}\right).
\label{gapeq}
\end{equation}
Here $1/\tau_{\rm pb} = 1/\tau_{\rm pb}^{\rm so} + 1/\tau_{\rm
pb}^{\rm si}$ where $1/\tau_{\rm pb}^{\rm so}$ and $1/\tau_{\rm
pb}^{\rm si}$ are the pair-breaking rates from spin-orbit and
spin-independent scattering, respectively, and are given by
\widetext
\FL
\begin{equation}
{1\over{\tau_{\rm pb}^{\rm so}}}={\pi\over 2} \int
{d\theta_{\bf k}\over {2\pi}}
N(\theta_{\bf k})
\int{d\theta_{{\bf k}^\prime}\over {2\pi}}
N(\theta_{{\bf k}^\prime})
|v_{{\bf k},{\bf k}^\prime}|^2
(\phi_d({\bf k})-\phi_d({\bf k}^\prime))^2\label{taupb}
\end{equation}
\narrowtext\noindent
and a similar expression with $|v_{{\bf k},{\bf k}^\prime}|^2$
replaced by $|w_{{\bf k},{\bf k}^\prime}|^2$ for $1/\tau_{\rm pb}^{\rm
si}$.  Equation (\ref{gapeq}) is precisely the same as the equation
for the suppressed $T_c$ of a conventional $s$-wave superconductor in
the presence of magnetic impurities \cite{maki}. The standard analysis
then shows that $T_c$ is reduced to zero when $1/\tau_{\rm pb} = \pi
T_{c0}/2\gamma \simeq 0.88 T_{c0}$, where $T_{c0}$ is the transition
temperature when $1/\tau_{\rm pb} = 0$ and the reduced transition
temperature is $T_c$ \cite{maki}.

To calculate $1/\tau_{\rm pb}$ it is necessary to know the correlation
functions (\ref{ccf}) and (\ref{fcf}) which characterize the
structural disorder.  The LTT and {\it Pccn} phases of the
(La,Nd)-(Ba,Sr)-Cu-O system are stabilized by randomly placed Nd or Ba
ions at La sites.  It is plausible that these randomly placed ions
alter the local tilting environment so that the average structure is
well defined, but locally the CuO$_6$ octahedra tilt about random
axes.  A simple model structure which may capture the essence of this
type of disorder is one in which Cu0$_6$ octahedra tilt coherently on
length scales less than a structural coherence length, $\xi_s$, while
on longer length scales the structure is completely disordered.  In
the presence of such disorder the function $C({\bf q})$ is peaked at
${\bf q} ={\bf Q}$ and has a width $\Delta q \simeq 1/\xi_s$.  For the
calculations presented below we use $C_{ab}({\bf q}) \propto
\exp(-2\xi_s^2 ({\bf q}-{\bf Q})^2)\delta_{a,b}$ where the
normalization is fixed by the requirement that the integral of $C({\bf
q})$ over the Brillouin zone must equal the mean square dispacement
angle $\theta_0^2$.  To allow a comparison of the relative importance
of spin-orbit and spin-independent scattering it is further assumed
that the disorder is Gaussian so that $F_{ab}({\bf q}) =
(2\pi)^{-2}\int d^2p C({\bf q}+{\bf p})C(-{\bf p})
\delta_{a,b}$.

First consider uncorrelated disorder ($\xi_s \rightarrow 0$).
Performing the integral (\ref{taupb}) for this case using a
nearest-neighbor tight-binding band structure and taking a chemical
potential of $\mu = -0.15 t$ yields $1/\tau_{\rm pb}^{\rm so}
\simeq 3.7 t \nu^2 \theta_0^2$ and $1/\tau_{\rm pb}^{\rm si}
\simeq 1.3 t \rho^2 \theta_0^4$.  These pair-breaking rates illustrate
the importance of including spin-orbit coupling when treating electron
scattering from structural disorder in a Cu-O layer.  Spin-orbit
scattering gives a pair-breaking rate which is quadratic in the root
mean square displacement, while spin-independent scattering gives a
rate which is quartic.  However, the spin-orbit scattering rate also
contains a factor of $\nu^2 \sim 4\times10^{-2}$ and so, for $\theta_0
\sim 0.1$, in the presence of uncorrelated disorder, spin-orbit and
spin-independent scattering are roughly of equal strength.

When $\xi_s$ is increased, pair-breaking from spin-orbit and
spin-independent scattering are no longer comparable in magnitude.
Figure \ref{pairbreakingtemp} shows the `pair-breaking temperature'
$T_{\rm pb} \equiv 2\gamma/\pi\tau_{\rm pb}$ due to spin-orbit and
spin-independent scattering, plotted vs.~$\xi_s$ (in units of the
lattice spacing) for $\mu = -0.15t$, $\gamma = 0.2$ and $\rho = 3.4$
(this value of $\rho$ is chosen for convenience so that $1/\tau_{\rm
pb}^{\rm so} = 1/\tau_{\rm pb}^{\rm si}$ when $\xi_s=0$).  Any
superconductor with a $d_{x^2-y^2}$ gap function which, in the absence
of disorder, has a critical temperature $T_{c0} < T_{\rm pb}$ will
have its $T_c$ reduced to zero when the pair-breaking lifetime is
$\tau_{\rm pb}$.  As $\xi_s$ increases pair-breaking from
spin-independent scattering is suppressed and pair-breaking from
spin-orbit scattering is enhanced.  Note that for some parameters the
pair-breaking from spin-orbit scattering can be strong enough to
reduce to zero the $T_c$ of a superconductor with $T_{c0} \sim 30$K.

The reason for this enhancement is illustrated in
Fig.~\ref{brillouinzone}.  This figure shows the Fermi surface for a
nearest-neighbor tight-binding band at 10\% doping, a typical ${\bf
k}$ point on that Fermi surface, and the region in momentum space
containing those points ${\bf k}^\prime$ for which the spin-orbit
scattering matrix element $|v_{{\bf k},{\bf k}^\prime}|^2$ is large.
This region is centered at ${\bf k}+{\bf Q}$ and has linear dimension
$\xi_s^{-1}$.  When $\xi_s$ is large the region does not touch the
Fermi surface and spin-orbit scattering is not an effective
pair-breaker.  As $\xi_s$ decreases the region grows, at some point
touches the Fermi surface, and electrons begin to be strongly
scattered.  This `focussed' large momentum scattering transfers
electrons primarily between regions of the Fermi surface where a
$d_{x^2-y^2}$ gap has different parities.  As a result the anomalous
and normal self-energy contributions to (\ref{taupb}) add coherently
rather than cancel as they do for a conventional $s$-wave
superconductor \cite{maki}.  This is what gives rise to strong pair
breaking.

To summarize, the scattering of electrons in a Cu-O plane from O
displacements perpendicular to that plane has been investigated.  The
leading order source of this scattering, in powers of displacement
size, is spin-orbit coupling.  Within the RPA, the spin-orbit
scattering vertex is not enhanced by spin fluctuations, unlike
scattering from an impurity spin.  Also, for a superconductor with a
$d_{x^2-y^2}$ gap function, spin-orbit scattering can be a strong pair
breaker, particularly in a structurally disordered phase in which
CuO$_6$ octahedra tilt coherently on small length scales, but are
completely disordered on longer length scales.  It is possible that
the LTT and {\it Pccn} phases of the La-Nd-Sr-Cu-O and La-Ba-Cu-O
systems have more structural disorder than the LTO phase, and that
pair-breaking effects such as those discussed here are responsible for
the observed suppression of superconductivity in these phases.  If
this is the case, then these observations are strongly suggestive of
unconventional pairing and support the $d$-wave hypothesis of
high-$T_c$ superconductivity.

I would like to acknowledge useful discussions with B. B\"uchner, R.
Hlubina, A. Kampf, and in particular T.M. Rice and F.C. Zhang.  This
work was supported by the Swiss National Fund.
%

%
\figure{
Diagrammatic representation of the self-consistent equation for the
RPA enhancement of a scattering vertex for a generic spin-flip
scatterer (only the diagrams for the $\sigma^{-}$ component are
shown).  For spin impurity scattering ($X_{\bf q} = S_{\bf q}$) the
solution to this equation gives a typical RPA enhancement factor of
$(1-U\chi_0({\bf q}))^{-1}$ while for spin-orbit scattering ($X_{\bf
q} = \theta_{\bf q}$) there is no such enhancement.
\label{diagrams}}

\figure{
Pair-breaking temperature, $T_{\rm pb} = 2\gamma/\pi\tau_{\rm pb}$,
characterizing the pair-breaking effectiveness of spin-orbit
scattering (solid line) and spin-independent scattering (dashed line)
for a $d_{x^2-y^2}$ superconductor in a structurally disordered phase
(see text), plotted as a function of the structural coherence length
of that phase, $\xi_s$.  The parameter values used are $\mu = -0.15t$,
$\theta_0 = 0.1$, $\nu = 0.2$ and $\rho \simeq 3.4$.  The value of
$\rho$ has been chosen so that for uncorrelated disorder ($\xi_s = 0$)
spin-orbit and spin-independent scattering are equally effective pair
breakers. The enhancement of $T_{\rm pb}$ for spin-orbit scattering is
due to the focussed scattering across the Fermi surface shown in
Fig.~\ref{brillouinzone}.
\label{pairbreakingtemp}}

\figure{
Brillouin zone for a 2D square lattice showing the tight-binding Fermi
surface for 10\% doping.  The zone is divided into four quadrants
marked $+$ or $-$ according to the parity of a $d_{x^2-y^2}$ gap.  A
typical point on the Fermi surface, ${\bf k}$ is marked with a black
dot, as well as the point ${\bf k}+(\pi,\pi)$ where the initial
electron would be Bragg scattered by a coherent staggered distortion.
The circle surrounding the shifted point contains the region within
which elastic spin-orbit scattering is strongest in a structurally
disordered phase with a structural correlation length $\xi_{s}$.
Because electrons are scattered most strongly across the Fermi surface
from regions where the gap is positive to regions where it is negative
this type of scattering is a particularly effective pair-breaker for a
$d_{x^2-y^2}$ superconductor.\label{brillouinzone}}

\end{document}